\documentclass[pra,aps,preprint,showpacs]{revtex4}
\newcommand{\ket}[1]{|#1\rangle}
\newcommand{\bra}[1]{\langle #1|}

\begin{document}
\title{Sufficiency Criterion for the Validity of the Adiabatic Approximation}
\author{D. M. Tong$^{1,2}$
\footnote{Electronic address: phytdm@nus.edu.sg}
, K. Singh$^{1}$, L. C. Kwek$^{1,3}$,  and C. H. Oh$^1$\footnote{Electronic address: phyohch@nus.edu.sg}}
\affiliation{
$^1$
Department of Physics, National University of
Singapore, 10 Kent Ridge Crescent, Singapore 119260, Singapore \\
$^2$Department of Physics, Shandong Normal University, Jinan 250100, China\\
$^3$National Institute of Education, Nanyang Technological
University, 1 Nanyang Walk, Singapore 639798, Singapore}
\date{\today}
\begin{abstract}
We examine the quantitative condition which has been widely used as a criterion for the adiabatic approximation but was recently found insufficient. Our results indicate that the usual quantitative condition is sufficient for a special class of quantum mechanical systems. For general systems, it may not be sufficient, but it along with additional conditions is sufficient. The usual quantitative condition and the additional conditions constitute a general criterion for the validity of the adiabatic approximation, which is applicable to all $N-$dimensional quantum systems. Moreover, we illustrate the use of the general quantitative criterion in some physical models. 
\end{abstract}
\maketitle
\date{\today}

The adiabatic theorem reads that if a quantum system with a time-dependent nondegenerate Hamiltonian $H(t)$ is initially in the $n$-th instantaneous eigenstate of $H(0)$, and if $H(t)$ evolves slowly enough, then the state of the system  at time $t$ will remain in the $n$-th instantaneous eigenstate of $H(t)$ up to a multiplicative phase factor. The theorem is a useful tool \cite{Born,Schwinger,Schiff,Bohm,Kato,Messiah,Landau} but its practical application relies on the criterion of the ``slowness" required by the theorem. In the literature, the ``slowness" is usually encoded by the quantitative condition, 
\begin{eqnarray}
\left|\frac{\langle{E_n(t)}\ket{\dot
E_m(t)}}{E_n(t)-E_m(t)}\right| \ll 1,~m\neq n,~ t\in[0,\tau]\label{cons}
\end{eqnarray}
%or more strictly, $\max_{t\in[0,\tau]}|\langle{E_n(t)}\ket{\dot E_m(t)}|\ll \min_{t\in[0,\tau]}|E_n(t)-E_m(t)|$
where $E_m(t)$ and $\ket{E_m(t)}$ are the eigenvalues and eigenstates of $H(t)$, and $\tau$ is the total evolution time. 
This quantitative condition had been deemed to be a sufficiency criterion, but it was recently found insufficient. In order to resolve the counterexample raised in \cite{Marzlin}, we showed that fulfilling only the quantitative condition cannot guarantee the validity of the adiabatic approximation\cite{Tong2}. While this explains the counterexample in \cite{Marzlin}, it raises the obvious question: In what situations would the criterion be sufficient and more importantly how can it be extended to general cases? There have been some attempts to address different aspects of this problem\cite{Mackenzie, Vertesi, Larson, Ye}, but it remains largely unresolved.

In order to resolve the problem, we trace the quantitative condition to its source, and we note that the condition has never been convincingly proven. It looks likely that the condition was first derived from some special quantum systems and it was extended to systems beyond its range of applicability. For instance, in  Refs. \cite{Bohm,Schiff}, condition (\ref{cons}) was obtained by assuming a first order approximation and by requiring both $E_m(t)-E_n(t)$ and $\langle E_m(t)\ket{\dot E_n(t)}$ to be constants. However, such a proof is not convincing because a first order approximation may be taken as a good approximation of the exact value only if all the higher order corrections are proven to be much smaller. We note that this is not the case here. This has also been pointed out in Refs. \cite{Mackenzie,Vertesi}. Therefore, even if the condition is sufficient for the special quantum systems, a convincing proof is still necessary. In any case, the sufficiency criterion for general systems is grossly lacking. It should be emphasized that the lack of a sufficiency criterion weakens the applicability of the adiabatic theorem. In the present paper, we address this sufficiency criterion issue. Firstly, we furnish a new proof to show that the quantitative condition is indeed a sufficiency criterion for the adiabatic approximation in the quantum systems which satisfy the requirement of both $E_m(t)-E_n(t)$ and $\langle E_m(t)\ket{\dot E_n(t)}$ being constant. Secondly, to extend its validity, we show that the the quantitative condition along with some additional conditions is sufficiency for general systems. Thus, the usual quantitative condition and the additional conditions constitute a general criterion for the adiabatic approximation, which is applicable to all $N-$dimensional quantum systems. Moreover, we illustrate the use of the general criterion in some physical models. 

Let us consider an $N$-dimensional quantum system with the Hamiltonian $H(t)$. The instantaneous nondegenerate eigenvalues and orthonormal eigenstates of $H(t)$, denoted as $E_m(t)$ and $\ket{E_m(t)}$ respectively, are defined by
\begin{eqnarray}       
H(t)\ket{E_m(t)}=E_m(t)\ket{E_m(t)},~m=1,\ldots,N.
\label{H}       
\end{eqnarray}
$\ket{E_m(t)}$ is determined by Eq. (\ref{H}) up to a phase factor. Hereafter, we choose it such that $\langle E_m(t)\ket{\dot E_m(t)}=0$\cite{Tong8}.  
If we assume that the system is initially in the $n-$th eigenstate $\ket{\psi(0)}=\ket{E_n(0)}$, then its state at time $t$,  $\ket{\psi(t)}$, is dictated by the Schr\"odinger equation
\begin{eqnarray}
i\frac{d}{dt}\ket{\psi(t)}=H(t)\ket{\psi(t)}.
\label{schr}
\end{eqnarray}
In the basis $\{\ket{E_m(t)}\}$, $\ket{\psi(t)}$ can be expanded as  
\begin{eqnarray}
\ket{\psi(t)}=\sum_m c_m(t)e^{-i\int_0^tE_m(t')dt'}\ket{E_m(t)},
\label{psit}
\end{eqnarray}
where $c_m(t)=e^{i\int_0^tE_m(t')dt'}\bra{E_m(t)}\psi(t)\rangle$ are the time-dependent coefficients. Substituting it into the Schr\"odinger equation, we obtain
 \begin{eqnarray}
\frac{dc_m(t)}{dt}+\sum_{l\neq m}\langle E_m\ket{\dot
E_l}e^{i\int_0^{t}\omega_{ml}dt'} c_l(t)=0, \label{dotc}
\end{eqnarray}
which leads to 
\begin{eqnarray}
c_m(t)=\delta_{mn}-\sum_{l\neq m}\int_0^t\langle E_m\ket{\dot E_l}
e^{i\int_0^{t'}\omega_{ml}dt''}c_l(t')dt', \label{cmt}
\end{eqnarray}
where $E_m\equiv E_m(t)$, $\ket{E_m}\equiv \ket{E_m(t)}$, $\omega_{ml}\equiv E_m(t)-E_l(t)$, and $m=1,2,...,N$.  Here, $n$
 in Eq. (\ref{cmt}) is the index of the initial state $\ket{E_n(0)}$.  We want to ascertain the criterion under which the adiabatic approximation is valid,  {\sl i.e.}, the condition(s) for which the fidelity $F=\left|\langle E_n(t)\ket{\psi(t)}\right|=\left|c_n(t)\right|\approx 1$. 

We first discuss the quantum systems for which both $\omega_{ml}$ and $\langle E_m\ket{\dot E_l}$ are constants\cite{Tong6}. In this case, from Eq. (\ref{cmt}), we have, by partial integration, 
\begin{eqnarray}
c_n(t)=1+i\sum_{m\neq n}\frac{\langle E_n\ket{\dot E_m}}{\omega_{nm}}\left( e^{i\omega_{nm}t}c_m(t)-\int_0^t e^{i\omega_{nm}t'}\dot c_m(t')dt'\right).
\label{cnt}
\end{eqnarray}
Substituting $\dot c_m(t)$ from Eq. (\ref{dotc}) into the above equation, we obtain
\begin{eqnarray}
c_n(t)&=&1+i\sum_{m\neq n}\frac{\langle E_n\ket{\dot E_m}}{\omega_{nm}} e^{i\omega_{nm}t}c_m(t)\nonumber\\ &&+i\sum_{m\neq n}\sum_{l\neq m}\frac{\langle E_n\ket{\dot E_m}}{\omega_{nm}}\langle E_m\ket{\dot E_l} \int_0^te^{i\omega_{nl}t'} c_l(t')dt'.
\label{cnt3}
\end{eqnarray}
Noting that $\left|c_m(t)\right|\leq 1$, we have from Eq. (\ref{cnt3})
\begin{eqnarray}
1-\left|c_n(t)\right|\leq\sum_{m\neq n}\left|\frac{\langle E_n\ket{\dot E_m}}{\omega_{nm}} \right|\left(1
+\sum_{l\neq m}\left|\langle E_m\ket{\dot E_l}\right|\left|I_{l}\right|\right)
 ,\label{ccn}
\end{eqnarray}
where $I_{l}=\int_0^te^{i\omega_{nl}t'} c_l(t')dt'$.

Clearly, if the integral $I_{l}$ is bounded by a finite number, the quantitative condition (\ref{cons}) can sufficiently guarantee that $1-\left|c_n(t)\right|\ll 1$.  We now show that this is indeed the case. 
To this end, by letting $\overline c_l(t)=c_l(t) e^{i\omega_{nl}t}$ and substituting it into Eq. (\ref{dotc}), we have 
\begin{eqnarray}
\frac{d \overline{c}_m(t)}{dt}-i\omega_{nm}\overline{c}_m(t) +\sum_{l\neq m}\langle E_m\ket{\dot
E_l} \overline{c}_l(t)=0. \label{a3}
\end{eqnarray}
Since Eq. (\ref{a3}) is a system of differential equations, the general solution of $\overline c_{m}(t)$ comprise $N$ special solutions in the form $a_{m}e^{i\lambda t}$, where $a_m$ and $\lambda$ are time-independent constants. They are determined by the equations,
\begin{eqnarray}
\left(\omega_{nm}-\lambda\right)a_m+i\sum_{l\neq m}\langle E_m\ket{\dot
E_l} a_l=0, \label{a4}
\end{eqnarray}
where $m=1,2,...,N.$ Solving Eq. (\ref{a4}), we may obtain $\lambda=\lambda_1,~\lambda_2,...,\lambda_N$, where $\lambda_j$ are nonzero real numbers\cite{Tong3}. For each $\lambda_j$, there is a solution $\overline c_{mj}(t) = a_{mj}e^{i\lambda_j t}$. All the $N$ independent solutions lead to the general solution $\overline c_m(t)=\sum_{j=1}^Np_j a_{mj}e^{i\lambda_jt}$, where the coefficients $p_j$ are determined by the initial conditions $\overline c_m(0)=\delta_{mn}$. 
Then, we have $|I_l|=\left|\int_0^t \overline c_l(t') dt'\right|=\left|\sum_{j=1}^N\frac{p_j a_{mj}}{i\lambda_j}\left(e^{i\lambda_jt}-1\right)\right|\leq 2\sum_{j=1}^N\left|\frac{p_j a_{mj}}{\lambda_j}\right|,$ where the latter term is a finite number, independent of time $t$. This completes the proof that the quantitative condition (\ref{cons}) is a sufficiency criterion for the quantum systems which satisfy the requirement that both $\omega_{ml}$ and $\langle E_m\ket{\dot E_l}$ are constants. 

We now discuss general quantum systems. Let us return to Eq. (\ref{cmt}). We have, by partial integration, 
\begin{eqnarray}
c_n(t)
&=&1+i\sum_{m\neq n}\frac{\langle E_n\ket{\dot E_m}}{\omega_{nm}} e^{i\int_0^{t}\omega_{nm}dt'}c_m(t)\nonumber\\ &&-i\sum_{m\neq n}\int_0^t\left(\frac{\langle E_n\ket{\dot E_m}}{\omega_{nm}}\right)' e^{i\int_0^{t'}\omega_{nm}dt''}c_m(t')dt'\nonumber\\ &&-i\sum_{m\neq n}\int_0^t\frac{\langle E_n\ket{\dot E_m}}{\omega_{nm}} e^{i\int_0^{t'}\omega_{nm}dt''}\dot c_m(t')dt'.
\label{cntb}
\end{eqnarray}
Substituting Eq. (\ref{dotc}) into Eq. (\ref{cntb}), we have
\begin{eqnarray}
c_n(t)&=&1+i\sum_{m\neq n}\frac{\langle E_n\ket{\dot E_m}}{\omega_{nm}} e^{i\int_0^{t}\omega_{nm}dt'}c_m(t)\nonumber\\&&-i\sum_{m\neq n}\int_0^t\left(\frac{\langle E_n\ket{\dot E_m}}{\omega_{nm}}\right)' e^{i\int_0^{t'}\omega_{nm}dt''}c_m(t')dt' \nonumber\\&&+i\sum_{m\neq n}\sum_{l\neq m}\int_0^t\frac{\langle E_n\ket{\dot E_m}}{\omega_{nm}}\langle E_m\ket{\dot E_l} e^{i\int_0^{t'}\omega_{nl}dt''} c_l(t')dt'.
\label{cnt2b}
\end{eqnarray}
In the general case, although it is difficult to estimate exactly the values of the integrals in Eq. (\ref{cnt2b}) as we did in the above special case, it is still possible to obtain bounds on the integrals, which will lead to the sufficiency criterion. Noting that $\left| c_m(t)\right|\leq 1$ and $\left|e^{i\int_0^{t}\omega_{nm}dt'}\right|=1$, we have 
\begin{eqnarray}
\left|\int_0^t\left(\frac{\langle E_n\ket{\dot E_m}}{\omega_{nm}}\right)' e^{i\int_0^{t'}\omega_{nm}dt''}c_m(t')dt'\right|\leq \int_0^t\left|\left(\frac{\langle E_n\ket{\dot E_m}}{\omega_{nm}}\right)'\right|dt',
\label{ca61}
\end{eqnarray}
and
\begin{eqnarray}
\left|\int_0^t\frac{\langle E_n\ket{\dot E_m}}{\omega_{nm}}\langle E_m\ket{\dot E_l} e^{i\int_0^{t'}\omega_{nl}dt''} c_l(t')dt'\right|\leq\int_0^t\left|\frac{\langle E_n\ket{\dot E_m}}{\omega_{nm}}\right|\left|\langle E_m\ket{\dot E_l}\right|dt'\label{ca62}.
\end{eqnarray}
From Eqs. (\ref{cnt2b}), (\ref{ca61}), and (\ref{ca62}), we obtain
\begin{eqnarray}
1-\left|c_n(t)\right|\leq &&\sum_{m\neq n}\left|\frac{\langle E_n\ket{\dot E_m}}{\omega_{nm}}\right|
+\sum_{m\neq n}\int_0^t\left|\left(\frac{\langle E_n\ket{\dot E_m}}{\omega_{nm}}\right)'\right|dt'
\nonumber\\
&&+\sum_{m\neq n}\sum_{l\neq m}\int_0^t\left|\frac{\langle E_n\ket{\dot E_m}}{\omega_{nm}}\right|\left|\langle E_m\ket{\dot E_l}\right|dt'.
\label{ca7}
\end{eqnarray}
Since the sums on the right hand side of expression (\ref{ca7}) are finite terms, the approximation $1-\left|c_n(t)\right|\ll 1$ is guaranteed if each of the terms is small. This is met if the following conditions 
\begin{eqnarray}
(A)~~&&\left|\frac{\langle E_n(t)\ket{\dot E_m(t)}}{E_n(t)-E_m(t)}\right| \ll 1,~~t\in[0,\tau],\label{1}\\
(B)~~&&\int_0^\tau\left|\left(\frac{\langle E_n(t)\ket{\dot E_m(t)}}{E_n(t)-E_m(t)}\right)'\right|dt\ll 1,\label{ca8b}\\
(C)~~&&\int_0^\tau\left|\frac{\langle E_n(t)\ket{\dot E_m(t)}}{E_n(t)-E_m(t)}\right|\left|\langle E_m(t)\ket{\dot E_l(t)}\right|dt\ll 1,\label{ca8c}
\end{eqnarray}
are satisfied, where $m\neq n$ and $\tau$ is the total evolution time for which the adiabatic approximation is valid, $t\in[0,\tau]$. Expression $(A)$ is just the well-known quantitative condition (\ref{cons}), and expressions $(B)$ and $(C)$ are two additional conditions, which set a bound on the total evolution time. Physically, if a quantum system satisfying condition $(A)$ is initially in its $n-$th eigenstate, it will remain close to its $n-$th instantaneous eigenstate during the initial short time but it may deviate from its instantaneous eigenstate at a later time. The additional conditions provide a time scale, $\tau$, for which the state remains close to the instantaneous eigenstate. $\tau$ can be obtained after calculating the integrals in Eqs. (\ref{ca8b}) and (\ref{ca8c}). In some special cases, the integrals may be easily evaluated.  For instance, if  $E_n(t)-E_m(t)$ is a monotonic function of $t$, the integral in Eq. (\ref{ca8b}) can be carried out. In some other cases, we may not be able to evaluate the integrals analytically. In such instance, we may simplify the conditions by appealing to estimations.  In any case, the stronger expressions   
\begin{eqnarray}
(b)~~&&\left|\left(\frac{\langle E_n(t)\ket{\dot E_m(t)}}{E_n(t)-E_m(t)}\right)'\right|_M\tau\ll 1,\label{a8b}\\
(c)~~&&\left|\frac{\langle E_n(t)\ket{\dot E_m(t)}}{E_n(t)-E_m(t)}\right|_M\left|\langle E_m\ket{\dot E_l}\right|_M\tau\ll 1, 
\label{a8c}
\end{eqnarray}
can always cover conditions $(B)$ and $(C)$ respectively, where $|f(t)|_M$ means the maximal modulus of $f(t)$ for $t\in[0, \tau]$. 
As conditions $(b)$ and $(c)$ are stronger than $(B)$ and $(C)$, the latter should be preferentially used when possible.    

The quantitative condition $(A)$ ({\sl i.e.}(\ref{cons})) and the additional conditions $(B)$ and $(C)$ constitute a general quantitative criterion for the adiabatic approximation. The general criterion can sufficiently guarantee the validity of the approximation and it is applicable to all $N-$dimensional quantum systems. That is, if a quantum system, initially in the eigenstate $\ket{E_n(0)}$, fulfills the general criterion,  it will remain, with high probability, in the $n$-th instantaneous eigenstate $\ket{E_n(t)}$ up to a phase factor. The fidelity between the approximate state and the exact state may be estimated by Eq. (\ref{ca7}).  It is interesting to use this criterion to reexamine the counterexample furnished with two related Hamiltonians $H^a(t)=i\dot U(t)U^+(t)$ and $H^b(t)=i\dot U^+(t)U(t)$ in \cite{Marzlin,Tong2}. Suppose the eigenstate $\ket{E^i_m}$ of $H^i(t)~(i=a,b)$ has been properly chosen such as $\langle E^i_m\ket{\dot E^i_m}=0$. We may have the relation, $\langle E^b_n\ket{\dot E^b_m}= e^{i\int_0^t(E^a_n-E^a_m)dt'}\langle E^a_n\ket{\dot E^a_m}$, which result in that condition $(B)$ cannot be satisfied for $H^b(t)$, in general, if it is satisfied for $H^a(t)$. Hence, the counterexample is ruled out from the adiabatic systems. We now apply the general criterion to some quantum systems.

Firstly, we specialize the general criterion for the quantum systems in which $E_n(t)-E_m(t)$ is a monotonic function of $t$. Many interesting adiabatic systems may belong to this class. In this case, we have   
\begin{eqnarray}
 \int_0^\tau\left|\left(\frac{\langle E_n\ket{\dot E_m}}{\omega_{nm}}\right)' \right|dt\leq \left|\frac{\langle E_n\ket{\dot E_m}}{\omega_{nm}}\right|_M\left|\ln\frac{\omega_{nm}(\tau)}{\omega_{nm}(0)}\right| +\left|\frac{\langle E_n\ket{\dot E_m}'}{\omega_{nm}}\right|_M\tau.
\label{imnc}
\end{eqnarray}
Since $|\ln\omega_{nm}(\tau)/\omega_{nm}(0)|$ is a finite number and hence the first term is small under condition $(A)$, the additional condition $(B)$ can be written as 
\begin{eqnarray}
(B_1)~~~~\left|\frac{\langle E_n\ket{\dot E_m}'}{\omega_{nm}}\right|_M\tau\ll 1. 
\label{a8c3}
\end{eqnarray}
As we cannot simplify the integral in (\ref{ca8c}) here, we use the stronger expression $(c)$. 
Therefore, conditions $(A)$, $(B_1)$ and (c) constitute the sufficiency criterion. Since in real physical experiments the total evolution time $\tau$ is usually finite, condition $(c)$ is automatically ensured by condition $(A)$. Hence, in this case, we may take $(A)$ and $(B_1)$ as the adiabatic criterion. 

Secondly, we consider a quantum system defined by the parameterized Hamiltonian $H(s)$, where $s=t/T,~t\in[0,T]$.
%, with the $N$ nondegenerate eigenvalues $E_m(s)$ and eigenstates $\ket{E_m(s)}$. 
The well-known proofs of adiabatic theorem given in Refs. \cite{Kato, Messiah} were carried out by using such a Hamiltonian. We now apply the general criterion to the system. By substituting $t=Ts$ into conditions $(A)$, $(B)$, and $(C)$, we obtain  
\begin{eqnarray}
&&\left|\frac{\langle E_n\ket{\dot E_m}}{E_n-E_m}\right|_M =  \frac{1}{T}\left|\frac{\langle E_n(s)\ket{\dot E_m(s)}}{E_n(s)-E_m(s)}\right|_M, \nonumber\\
&&\int_0^T\left|\left(\frac{\langle E_n\ket{\dot E_m}}{E_n-E_m}\right)'\right|dt\leq \frac{1}{T}\left|\frac{\langle E_n(s)\ket{\dot E_m(s)}}{E_n(s)-E_m(s)}\right|_M\left|\frac{\left(E_n(s)- E_m(s)\right)'}{E_n(s)-E_m(s)}\right|_M+\frac{1}{T}\left|\frac{\langle E_n(s)\ket{\dot E_m(s)}'}{E_n(s)-E_m(s)}\right|_M, \nonumber\\
&&\int_0^T\left|\frac{\langle E_n\ket{\dot E_m}}{E_n-E_m}\right|\left|\langle E_m\ket{\dot E_l}\right|dt\leq \frac{1}{T}\left|\frac{\langle E_n(s)\ket{\dot E_m(s)}}{E_n(s)-E_m(s)}\right|_M\left|\langle {E_m(s)}\ket{\dot E_l(s)}\right|_M.
\label{a16}
\end{eqnarray}
Since all the terms on the right of the above expressions can be arbitrarily small as $T$ becomes large, conditions $(A)$, $(B)$, and $(C)$ are met if $T$ is large enough. We then arrive at the conclusion that the adiabatic approximation is always valid for quantum systems defined by Hamiltonians of the forms  $H(\frac{t}{T})$ with $t\in[0,T]$, as long as $T$ is large enough. This conclusion agrees with the results in \cite{Kato,Messiah}. What is special about such Hamiltonians is that the criterion expressed by $(A)$, $(B)$, and $(C)$ can always be satisfied by choosing values of the parameter $T$. However, once the parameter $T$ is chosen, there is also a bound on total evolution time, $\tau=T$. The adiabatic approximation is valid if $t\in[0,T]$, and it may be invalid if $t$ is larger than $T$.   

Finally, we apply the general criterion to a concrete model to understand how the time constraint functions. Consider a spin-half particle in a rotating magnetic field, $H(t)=-\frac{\omega_0}{2}(\sigma_x\sin\theta\cos\omega t+\sigma_y\sin\theta\sin\omega t+\sigma_z\cos\theta)$.
For this model, we have $E_1-E_2=\omega_0$, $\langle E_1\ket{\dot E_2}=-\frac{i\omega}{2} \sin\theta e^{i\omega t\cos\theta}$. Substituting them into $(A)$, $(b)$, and $(c)$, we have
$\omega\sin\theta/\omega_0\ll 1$ , $(\omega\sin\theta/\omega_0)\cdot \omega \tau\cos\theta \ll 1$, and $(\omega\sin\theta/\omega_0)\cdot \omega \tau\sin\theta \ll 1$. It shows that, besides the usual condition, there is a time constraint, $t\in\left[0, ~k\frac{1}{\omega}\right]$, where $k$ is a certain number. The time constrain limits the total evolution time to a finite number of rotating periods of the magnetic field. Such a time limit is acceptable in physics, and it is also consistent with the geometric phase consideration. In Ref. \cite{Tong1}, it was shown that the geometric phase calculated by using the adiabatic approximation may differ appreciably from its exact value if the evolution time is too large. The difference reads $ \delta\gamma \simeq -\omega \tau\sin\theta \cdot\frac{\omega\sin\theta}{2(\omega_0+2\omega\cos\theta)}$. This implies that, in order to guarantee $\delta\gamma\ll 1$,  $\omega \tau$ must be finite. So, the application of the adiabatic approximation on geometric phase indicates that the time constraint is reasonable. 

In summary, we have examined the quantitative condition (\ref{cons}). Our results indicate that the usual quantitative condition (\ref{cons}) itself is a sufficiency criterion for the adiabatic approximation when it is applied to the quantum systems in which $(E_m-E_l)$ and $\langle E_m\ket{\dot E_l}$ are constants. It may not be a sufficiency criterion when it is applied to a general quantum system. To extend its validity to general systems, we have shown that the usual condition (\ref{cons}) and the additional conditions (\ref{ca8b}) and (\ref{ca8c}) constitute a general criterion for the adiabatic approximation. The general criterion can sufficiently guarantee the validity of the adiabatic approximation and it is applicable to all $N-$dimensional quantum systems. We have examined a few examples to illustrate its use. It should be noted that when the sufficiency conditions are used, the eigenstates $\ket{E_m}$ need to be properly chosen such that $\langle E_m\ket{\dot E_m}=0$.

\vskip 0.3 cm D. M. Tong acknowledges the useful discussions with  B. C. Sanders,  K.-P. Marzlin, and L.-A. Wu. This work was supported by NUS Research Grant No. R-144-000-189-305 and NSF of China No. 10675076. Tong also acknowledges the financial support from IQIS at University of Calgary when he visited there.


\begin{thebibliography}{99}
\bibitem{Tong8}If $\ket{E_m}$ does not satisfy $\langle E_m\ket{\dot E_m}= 0$, then it is replaced by $e^{-\int_0^t \langle E_m\ket{\dot E_m} dt'}\ket{E_m}$.
\bibitem{Tong6} One example of such systems is a spin-half particle in a rotating magnetic field with $\theta=\pi/2$. See the last page.%example in the present paper.
\bibitem{Tong3} Since $\lambda$ can be taken as the eigenvalues of the hermitian matrix defined by
$M_{ml}= \omega_{nm}\delta_{ml}+i\langle E_m(t)\ket{\dot E_l(t)}$, they must be real numbers. Besides, $\lambda=0$ is not a solution in general, otherwise, it leads to a trivial proof.
\end{thebibliography}
\end{document}